\documentclass[aps,prc,showpacs]{revtex4}
\usepackage{amsmath,epsfig}

\usepackage{graphicx}% Include figure files
\usepackage{dcolumn}% Align table columns on decimal point
\usepackage{bm}% bold math

\begin{document}
\title{Flow Coefficients and Jet Characteristics in Heavy Ion Collisions}

\author{\normalsize Sadhana Dash${}^1$, Dipak K. Mishra${}^2$, S. C. Phatak${}^1$ and P. K. Sahu${}^1$\\
\small ${}^1$Institute of Physics, Bhubaneswar 751 005, Orissa, India\\
\small ${}^2$GSI, D-64291 Darmstadt, Germany}

%%%%%%%%%%%%%%%%%%%%%%%%%%%%%%%%%%%%%%%%%%%%%%%%%%%%%%%%%%%%%%%%%%%%%%%%%%%%%%%%%

\begin{abstract}
\noindent Identifying jets in heavy ion collisions is of significant interest
since the properties of jets are expected to get modified because of
the formation of quark gluon plasma. The detection of jets is, however,
difficult because of large number of non-jet hadrons produced in the
collision process. In this work we propose a method of identifying
a jet and determining its transverse momentum by means of flow
analysis. This has been done an event-by-event basis.
\end{abstract}

\pacs{12.38.Mh, 24.85.+p, 25.75.-q}
\maketitle
\medskip
\noindent {\bf Keywords:} Fourier analysis, Relativistic heavy-ion collisions, collective flow, 
Jet properties and flow

\section{Introduction}

Identification of jets in heavy ion collisions is an important and
challenging problem for several reasons. First, the quenching of
jets has been proposed as one of the signatures of  the formation of
quark gluon plasma (QGP) \cite{gyulassy1,gyulassy2,gyulassy3}. It is
expected that when the leading parton, which eventually fragments into
jet particles, passes through a medium consisting of QGP it would loose
some or all of its energy, in a process which is analogous to the
loss of energy of a fast charged particle in the electromagnetic plasma.
The leading parton would also produce secondary quarks and gluons during
the passage \cite{gyulassy4}, resulting in the change of the profile of the
jet particles. Thus, the characteristics of jets produced in heavy ion
collisions would be different from those produced in hadron-hadron or
$e^+-e^-$ collisions. It has been argued that such a large modification
of jets would not occur when the parton passes through hot hadronic
gas. Jet quenching has already been observed by different
experiments at RHIC \cite{phenixwp,starwp}.
These studies are basically the correlation
studies between the energetic hadrons which are expected to be the
leading
hadrons produced in a jet. However, it would be interesting to detect jets
directly, as done in $e^+-e^-$ and hadron-hadron collisions or at
least identify them in heavy ion collisions.

The jet properties, like the number of particles in a jet and the
opening angle of the jet which are produced in elementary collisions,
have been well investigated  \cite{qcdjets1,qcdjets2}. Also there is
a good theoretical understanding of these in terms of perturbative
QCD and jet fragmentation functions \cite{qcdjets2,qcdjets3}. In
case of nucleus-nucleus collisions, identification of the jets using
the standard jet reconstruction algorithms such as cone or $k_T$
algorithms \cite{cone} is extremely difficult, if not impossible,
because of the large number of non-jet background particles produced
during the collision of two heavy ions. It is true that the jet
particles are produced in a narrow cone in $\eta$ ( the rapidity )
and $\phi$ ( the azimuthal angle ) and have large momenta. On the
other hand the background particles are distributed over a large
range of $\eta$ and spread more or less uniformly in $\phi$.
Nevertheless there are large number of non-jet particles in the jet
cone. This makes the removal of non-jet ( background ) particles
from the jet cone and identification of the jet particles difficult.
One may use momentum cutoff to filter out the bulk of the low
momentum background particles. Even then, it turns out that there
are  always sufficiently large number of background particles in the
jet cone for these algorithms to work.

Recently, we have developed a method for identification of jets in
heavy ion collisions \cite{sahu04}. The method is based on the fact
that the flow or Fourier coefficients for events containing jets have
a typical structure ( see later for the details ) which allows one to
identify the jet events, determine the jet opening angle and the
associated number of particles in the jet. It was also shown that when
there are two jets going back-to-back, the even flow coefficients are
significantly larger than the odd ones, which helps
in differentiating between the events having single jet and those having
two jets with (almost) opposite momenta.
In the present work we have
further extended this method by computing transverse momentum weighted
flow coefficients. Using these flow coefficients, we are able to estimate
the transverse momentum ($p_T$) of the jet as well as the jet opening
angle and the number of jet particles. We also find that when there are
two back-to-back jets present in the data, the flow coefficients for
even values of $m$ are larger and those for odd values of $m$ are close
to zero. This is particularly useful because the behavior of the flow
coefficients clearly differentiates between the events having back-to-back
jets ( that is no jet quenching ) and the events having single jet, in
which one of the fast parton is completely stopped.

We organize our work as follows: In Section 2, we briefly review the flow method of
characterizing jets and derive the results when the transverse momentum
weighted flow coefficients are computed. We then discuss
how these results, along with the results of the previous work
\cite{sahu04} can be used to determine the properties of the jet
which is followed by the analysis of the simulated data using our
method and determination of the jet characteristics in Section 3.
Section 4 concludes this study.

\section{Jet Identification From Flow Coefficients}

The flow coefficients are nothing but the Fourier coefficients of the
azimuthal distribution of particles produced in heavy ion collisions.
These are determined by doing Fourier analysis of the collisions data.
Thus, given a normalized distribution of particles,
$P(\phi)$ in azimuthal angle,
we can expand it in Fourier series \cite{voloshin}
\begin{equation}
P(\phi) =\frac{1}{2\pi} \sum_{m=0}^\infty 2  v_m \cos m ( \phi - \phi_m).
\end{equation}
The coefficients $v_m$'s are called flow coefficients
and $\phi_m$ is $m$-dependent angle \cite{defn}.
Generally, one expects $\phi_m$ to coincide with the reaction plane
from symmetry considerations. This is because the only preferred or special
plane in a collision is the reaction plane defined by the collision axis and the
impact parameter.
The flow coefficients $v_m$ are then given by
\begin{equation}\label{flow_def}
v_m = \int_0^{2\pi} d\phi P(\phi) \cos m(\phi - \phi_m),
\end{equation}
with $v_0=1$ because $P(\phi)$ is normalized.
The computation of $v_m$ from eq(\ref{flow_def}) requires the
knowledge of $\phi_m$. In the experiments, $\phi_m$ is not known
a priori and there are inaccuracies in the determination of $\phi_m$
from the data. It is therefore convenient to eliminate $\phi_m$ and
determine the flow coefficients by using two-particle correlation
method \cite{bevelac,bevelac1,ollitrault}. One then has
\begin{equation}
v_m^2 = \int d\phi_1 d\phi_2 P(\phi_1) P(\phi_2) \cos m (\phi_1 - \phi_2).
\end{equation}
In fact, it is advantageous to adopt the correlation method since it
eliminates the errors present in the determination of $\phi_m$ and
the errors arising from fluctuations due to  finite number of
detected particles. On the other hand, one of the disadvantages of
the correlation method is that the sign of the flow coefficients is
not determined. Our analysis, fortunately, does not depend on the
sign of $v_m$'s. It is well known that the flow coefficients for
$m=$1 and 2 give information about the early stages of the system
evolution in heavy ion collisions \cite{starflow, sorge, na49flow}.
These $v_1$ and $v_2$ are well established for a wide range of
energies\cite{sahu05,sahu02,sahu00,sahu98}. 
 Recently, there have
been some investigations on the physical interpretation of flow
coefficient for larger values of $m$
\cite{ollitrault1,kolb,lokhtin}. However, in our opinion, this
interpretation is not as appealing as the interpretation of $m=$1
and 2 in terms of asymmetry of collision and dynamical evolution of
colliding nuclei. In passing we would like to note that the
correlation method described above has been successfully applied for
flow analysis of different heavy ion collision experiments
\cite{starflow, na49flow}.

Our jet identification method is based on the fact that for particles
distributed uniformly in $\phi$, ( $P(\phi)=1/2\pi$ ) all the flow
coefficients ( except $v_0$ ) vanish. On the other hand,
for a $\delta$ function distribution all $v_m$'s are unity.
Thus if an event consists of a well-defined jet having a number of particles
produced in a small cone in $\phi$ which is embedded in a uniform background,
the flow coefficients would be abnormally large.
In the following we shall quantify this statement by considering
specific distribution functions for jet and background particles.
At this stage, we would like to point out the difference between
the usual calculations of flow coefficients, one is trying to deduce the
nature of the collective dynamics which is represented by the bulk of the
particle. In our calculation, on the other hand, we are interested in the
effect of a set of particles having common features (In this case, these
being emitted in a narrow cone in $\phi$) present in a large background,
on the flow coefficients.
The background particles may or may not have collective flow ( see later
for the case in which the event has a jet in the presence of collective
flow ).

The definition of the flow in eq(\ref{flow_def})
is not weighted
by any physical quantity. One often defines flow of a physical quantity
( say transverse momentum ) by weighting the averages by the
corresponding physical quantity \cite{voloshin}. Thus, the transverse
momentum flow $v_{m,p_T}$ is
\begin{equation}\label{flow_pT}
v_{m,p_{T}} = \int_0^{2\pi} \int_0^\infty d\phi dp_T p_T P(\phi, p_T) \cos m(\phi - \phi_m),
\end{equation}
We may eliminate $\phi_m$ and determine the transverse momentum flow as
\begin{eqnarray}
v_{m,p_T}^2=\int d\phi_1 d\phi_2 dp_{T1} dp_{T2}p_{T1}p_{T2}P(\phi_1, p_{T1})P(\phi_2, p_{T2})
\\ \nonumber
\times \cos m(\phi_1-\phi_2).
\end{eqnarray}

For a given particle distribution in an experiment, the flow coefficients
are determined by the following equations:
\begin{eqnarray}
v_m^2 & = & \frac{1}{N^2} \sum_{i,j} \cos m( \phi_i - \phi_j) \\
\mathrm{and} \nonumber \\
v_{m,p_T}^2 & = &  \frac{1}{N^2} \sum_{i,j} p_{T,i} p_{T,j} \cos m( \phi_i - \phi_j),
\end{eqnarray}
where $N$ is the number of particles in the event and $p_{T,i}$ is the
transverse momentum of $i^{th}$ particle. Note the self-correlation ($i=j$ terms) is
included in the expression above. Without self-correlation, $v_m^2$'s could be negative,
which is unphysical.

In the following we shall embed one or two jets in the background particles
in each event, then compute and
study the behavior of the flow coefficients. Considering a heavy ion
collision event in which $N_b$ number of background particles in the
event are distributed uniformly in azimuthal angle $\phi$ and their
transverse momentum distribution is $f_b(p_T)$. In addition there
are $N_j$ number of jet particles emitted in a jet cone of angle
$\Delta \phi$ centered at $\phi_0$ and their transverse momentum
distribution is $f(p_T)$. We shall assume that the angular distribution
is uniform between $\phi - \Delta \phi/2$ and $\phi + \Delta \phi/2$.
Thus, the probability distribution of $N = N_b + N_j$ particles can
be defined as
\begin{eqnarray}
P(\phi, p_T) = \frac{N_b}{N} P_b(\phi, p_T) + \frac{N_j}{N} P_j(\phi, p_T),
\end{eqnarray}
where
\begin{eqnarray*}
P_b(\phi, p_T) & = & \frac{1}{2 \pi}f_b(p_T) \;\;\; \; \; {\mathrm {for}} \; \;  0 < \phi < 2\pi \\
\mathrm{and}\nonumber\\
P_j(\phi, p_T) & = & \frac{1}{\Delta \phi} f(p_T),
\;\;\; \; \; {\mathrm {for}} \; \;
\phi_0 - \Delta \phi /2 < \phi < \phi_0 + \Delta \phi /2.\nonumber\\
\end{eqnarray*}
The transverse momentum distributions of background and jet particles
have been normalized to unity.
Thus $\int f(p_T) dp_T = \int f_b(p_T) dp_T = 1$.
For this distribution, the expression for flow coefficients $v_m$ is
%\begin{widetext}
\begin{eqnarray}\label{flow_coef}
v^2_m & = &  \int d\phi_1 d\phi_2 dp_{T1} dp_{T2}P(\phi_1, p_{T1})
P(\phi_2, p_{T2})\cos m (\phi_1 - \phi_2) \nonumber \\
 & = & \int d\phi_1 d\phi_2 dp_{T1} dp_{T2}P(\phi_1, p_{T1})
P(\phi_2, p_{T2})\nonumber\\
&\times & ( \cos( m \phi_1 )\cos( m \phi_2) + \sin( m \phi_1 )
\sin ( m \phi_2 )) \nonumber \\
 & = & \frac{N_j^2}{N^2}  \Big [ \int d\phi_1 dp_{T1} P_j(\phi_1, p_{T1})
\cos m (\phi_1 - \phi_0) \Big ]^2 \nonumber \\
& = & \frac{N_j^2}{N^2} \Big [ j_0 ( m \Delta \phi /2 ) \Big ]^2.
\end{eqnarray}
%\end{widetext}

The last line in eq(\ref{flow_coef}) follows from the fact that, by
definition, the two-particle distribution function factorizes into a
product of single particle distribution function and the jet
distribution function is symmetric about $\phi_0$. Further, the
background particles do not contribute since their distribution is
independent of the azimuthal angle and all trigonometric integrals
vanish. If we use the fact that the jet angle $\Delta \phi$ is small,
we can expand the cosine function in powers of $\phi_1 - \phi_0$, then
we get
\begin{eqnarray}\label{v_plot}
v^2_m = \frac{N_j^2}{N^2} \Big [ 1 - \frac {m^2 \Delta \phi^2}{12}  + {\cal O}
(  m^4  ) \Big ].
\label{fig:vmsqr}
\end{eqnarray}
The results in eq (\ref{fig:vmsqr}) has been obtained for a specific azimuthal
distribution of jet particles. For this distribution, the variance
$\sigma = \sqrt{(\phi^2_{av})-(\phi_{av})^2}=\frac{\Delta\phi}{\sqrt{12}}$.
In fact, for general azimuthal distribution function, we can
show that \cite{sahu04}
\begin{eqnarray}
v^2_m = \frac{N_j^2}{N^2} \Big [ 1 - m^2 \sigma^2 + {\cal O} ( m^4 )
\Big ],
\end{eqnarray}
where we assume that the jet particles are distributed
according to a probability distribution function which is symmetric
about $\phi_0$.
One important point to notice in the
expression above is that  the flow coefficients don't depend on the
distribution of the background particles, provided that these are
distributed uniformly in azimuthal angle. Also, the details of
transverse momentum dependence of the jet particles is integrated
out.

Following the same methodology, the expression for  $p_T$-weighted flow
coefficients can be determined. One can write
\begin{eqnarray}\label{vpt_plot}
v^2_{m,p_T} & = &  \frac{N_j^2 <p_T>^2}{N^2} \Big [ j_0 ( m \Delta \phi /2 ) \Big ]^2
\nonumber \\
 & = & \frac{N_j^2 <p_T>^2}{N^2} \Big [ 1 - \frac{ m^2 \Delta \phi^2}{12}
 + {\cal O} ( m^4  ) \Big  ] \nonumber \\
& = & \frac{N_j^2 <p_T>^2}{N^2} \Big [ 1 - m^2 \sigma^2 + {\cal O} ( m^4  )
\Big ],
\end{eqnarray}
where $<p_T> = \int dp_T p_T f(p_T)$ is the average transverse momentum
carried by a particle
in the jet. Thus $N_j<p_T>$ gives the total transverse momentum of the jet.

The expressions in eqs(\ref{v_plot}, \ref{vpt_plot}) clearly suggest
a method of obtaining jet properties from the flow coefficients. If one
plots $v^2_{m}$ and $v^2_{m, p_T}$ {\it vs} $m^2$, the points would lie
on a straight line. For $v^2_{m}$ {\it vs} $m^2$ the intercept of the
line on y-axis will yield the number of jet particles and from $v^2_{m, p_T}$
{\it vs} $m^2$ plot the intercept on y-axis will give the transverse
momentum of the jet ( since $N$, the total number of particles is known ). And
from the slope, one can determine the opening angle ($\sigma$ or $\Delta \phi$).
So, the procedure would be to fit a straight line to the computed flow
coefficients and determine the intercept and the slope. These in turn would
yield number of jet particles, jet transverse momentum and jet opening angle.
Note that for linear fit the fitting procedure is trivial, with the slope,
intercept and errors in these quantities being determined by algebraic
expressions. It should be obvious that the analysis described above
is necessarily event-by-event analysis\cite{morsch,morsch1}.

We now come to the case when there are two jets. In particular, we
shall consider two jets emitted at azimuthal angles $\phi$ and
$\phi + \pi$. This is the case of interest because we expect that a hard
parton scattering would produce such jets having equal and opposite
jet momenta. We expect that quenching of one of the jets would broaden
the other jet and/or produce more jet particles. Thus, the
characteristics of the two back-to-back jets would be different. Further,
in an extreme situation, the fast moving parton of one of the jets may
be completely absorbed in the medium leading to removal of one of the
jets. To consider such a situation, following particle distribution
function $P(\phi, p_T)$ is chosen.
%\begin{widetext}
\begin{eqnarray}
P(\phi, p_T) = \frac{N_b}{N} P_b(\phi, p_T)+\frac{N_{j1}}{N} P_{j1}(\phi, p_T)+\frac{N_{j2}}{N} P_{j2}(\phi, p_T),
\end{eqnarray}
where
\begin{eqnarray}
P_b(\phi, p_T) & = & \frac{1}{2 \pi}f_b(p_T) \;\;\; \; \; {\mathrm {for}} \; \;  0 < \phi < 2\pi \\
P_{j1}(\phi, p_T) & = & \frac{1}{\Delta \phi} f_1(p_T) \;\;\; \; \;
{\mathrm {for}} \; \;
\phi_0 - \Delta \phi_1 /2 < \phi <
\phi_0 + \Delta \phi_1 /2  \\
P_{j2}(\phi, p_T) & = & \frac{1}{\Delta \phi} f_2(p_T) \;\;\; \; \;
{\mathrm {for}} \; \;
\phi_0 - \Delta \phi_2 /2 < \phi + \pi <
\phi_0 + \Delta \phi_2 /2.
\end{eqnarray}
%\end{widetext}
Note that the number of jet particles, their opening angles and
%their
momentum distributions have been assumed to be different for the two
jets. The computation of the flow coefficients can be carried out in
the same fashion and the result is
\begin{eqnarray}
v^2_m & = &\frac{1}{N^2}\Big [N_{j1}j_0(\frac{m\Delta\phi_{j1}}{2})\\ \nonumber
& + &(-1)^m N_{j2}
j_0(\frac{m \Delta\phi_{j2}}{2})\Big ]^2
\end{eqnarray}
 and
\begin{eqnarray}
v^2_{m, p_T} & = & \frac{1}{N^2} \Big [ <p_{T,1}> j_0(\frac{m \Delta \phi_{j1}}{2})\cr
& &+ (-1)^m
<p_{T,2}> j_0(\frac{m \Delta \phi_{j2}}{2}) \Big ]^2.
\end{eqnarray}
The result can be understood as follows. As in the case of a single jet,
the background particles being uniformly distributed in azimuthal angle,
which do not contribute to the flow coefficients. But, for two-jet case,
there is nonzero contribution from the two particles in the same jet as
well as the two particles belonging to different jets. The former
corresponds to the square of the individual terms in the brackets of
expressions above. The latter corresponds to the crossed terms. The
factor of $(-1)^m$ in the expression above arises because the jet
angles of the two jets differ by $\pi$. As in the case of a single
jet, we may expand the Bessel function $j_0$ in the powers of its
argument. However this does not yield a simple enough formula which can be
used for determining the properties of the two jets. Further more, since
only a few values of flow
coefficients can be determined reliably, the determination of the
properties of both of the jets ( {\it i.e.} the number of particles in
each jet and the opening angles ) from the flow coefficients cannot be
done reliably. Nevertheless, we can deduce some qualitative conclusions
from these expressions. First, for two jet case, the flow coefficients
for even $m$ are significantly larger than those for odd $m$. Particularly,
if the two jets have similar opening angles and  numbers of jet particles,
the odd flow coefficients are expected to be close to zero and much
smaller than the even coefficients. On the other hand, if the opening
angle of one of the jets is broadened significantly, the corresponding
Bessel function would decrease rapidly with $m$ and the pattern would
look more like a single jet case. In reality, one may not get
back-to-back jets even in hadron-hadron collision. This is due to
the fact that the partons have internal motion within a hadron and sometimes a 
scattered parton may emit another hard parton, thus producing a more than 
two-jet like structure. In that case, the angle between two jets would not 
be $180 ^o$ but close to it and hence, the odd flow coefficients will not 
vanish but would still be small. We will discuss this in the next section, when
there will be two jets with background particles.

\section{Determination of Jet Properties}
In the preceding section we have shown that the flow coefficients have
a characteristically different structure when there are jets present in
the event.
In particular, we have shown that, for small enough jet opening angle
( $\Delta \phi$ ), the square of the flow coefficients, $v^2_m$, have
linear dependence on $m^2$. Further, we have shown that the intercept and
slope of a linear fit gives the information about the number of
particles in the jet, jet opening angle and the transverse momentum
of the jet. In this section we shall apply this method and describe
the results of such a
program. The calculation is performed as follows. For an event with
$N$ particles, we compute the flow coefficients
$v^2_m = \frac{\sum_{i,j} \cos(m(\phi_i - \phi_j))}{N^2}$. Fitting a
curve $b - c m^2$ to these values by minimizing $\chi^2$, we obtain
the number of jet particles ( $N_j^2 = b N^2$ ) and the jet opening
angle ( $\Delta \phi^2 = 12 c N^2/N_j^2 = 12c /b $ ). The fitting
procedure also gives the error in $N_j$ and $\Delta \phi$. A similar
calculation for $p_T$-weighted flow coefficients gives transverse
momentum of the jet ( $N_j <p_T>$ ) as well as $\Delta \phi$.

Before going on to the discussion of the results, let us first consider
the possible limitations of the method. One situation
where the method would fail is when the number of background particles
is large. This is because the expressions of the flow coefficients
( eqs(\ref{v_plot}, \ref{vpt_plot} ) ) have $N^2$ in the denominator.
So, when the number of background particles is much larger than the number
of jet particles, the flow coefficients would be numerically small. In that
case, there would be a large error in the determination of the intercept as
well as the slope from the flow coefficients. In fact, we can estimate
the limit an the number of background particle, above which the method
would fail.
Given that there are
$N_b$ background particles uniformly distributed in azimuthal angle, the
average number of background particles in the jet cone is
$N_c = \Delta \phi N_b / ( 2 \pi ) $. However,
the background particles are distributed statistically and the fluctuations
in the background particles in the jet cone is of the order of $\sqrt{N_c}$.
Thus the method would work so long as $\sqrt{N_c}$ is smaller than the number of
jet particles. This problem can be considerably reduced by removing small
transverse momentum particles (which are predominantly non-jet particles)
from the analysis. The method would also fail when $\Delta \phi$ is large.
This is because, in this case,  the power series expansion of the Bessel
function in eqs(\ref{v_plot}, \ref{vpt_plot}) is valid for few values
of $m$. Roughly speaking, $m \Delta \phi / \sqrt{12}$ should be less than 1
for the series
expansion of the Bessel function to work. This means that for typical
jet opening angle of 0.5 radians, $m$ should be restricted to 6 or less.
For larger $\Delta\phi$, $m$ is reduced further. Finally, we note that
$N$ cannot be too small either because for very
small $N$, there are large statistical fluctuations in the distribution
of the background particles. This gives rise to large spurious flow
coefficients from the background particles, thus affecting the fitting
procedure.

One may think that when the events have dynamical flow, such as elliptical
flow, this flow of the background particles will interfere with the flow
coefficients arising from the jet particles. Fortunately, this is not the
case because the values of the elliptic flow due to dynamics are not very
large. Typically, the dynamical  elliptical flow is observed to be of
the order of $10-15\%$. Since we are determining $v_m^2$, the contribution
of dynamical elliptic flow to $v_2^2$ is $\sim 0.01 - 0.02$. We shall
see later that in the presence of a jet, $v_m^2$'s are $\sim 0.1$. Thus,
the contribution of the dynamical flow much smaller than that due to a
jet and therefore the dynamical flow does not affect the determination of
jet properties very much.

To test the method we first generate events in which the background particles
are produced from HIJING event generator \cite{hijing} by switching off the jet 
production. The collision
energy is 5.5 TeV/nucleon in center of mass for $Pb + Pb$ collision.
This corresponds to the energy at LHC. Total of 3000 such events with
impact parameter between 3 and 7 fm have been generated.
Charged particles within one unit of rapidity at
mid-rapidity are considered for the analysis. We have then added a single
jet at a randomly chosen jet angle $\phi_0$ with $N_j$ number of jet
particles. The jet particles are assumed to be distributed uniformly
within a jet cone angle of $\Delta \phi$.
The normalized $p_T$-distribution function $f(p_T)=\alpha e^{-\alpha (p_T-\beta)}$
with $p_T$ between $\beta$  and $\infty$ and $\alpha = 1.2 $ GeV$^{-1}$ and $\beta = 1 $ GeV.
Thus $\int dp_T P(p_T) = 1$ and
$ < p_T > = \int dp_T p_T P(p_T) = \beta + 1/\alpha = 1.83$ GeV. The
analysis is done for the number of jet particles varying from 10 to 20,
the opening angle between $\pi/6$ and $\pi/4$ radians. The analyses are
done by using a $p_T$ cut of 0.5, 0.75 and 1 GeV for background particles.
The results are discussed below.

Further, we have also tested our method using HIJING event generator 
with default jet production in all events in the $Pb-Pb$
collision at 5.5 $TeV$/nucleon and between impact parameters 3 to 7 fm. 
This is done to check if our method can detect events having one and two 
jets successfully. Computations are done for $p_T=0.75$ GeV cuts and with and 
without $p_T$ weights. 
It may be noted that jet events with large energy and large number of jet 
particles are rare. So, not every event has dominant jet-like structure. 

\subsection{Only background particles}
\begin{figure}[htb]
\centerline{~{\psfig{figure=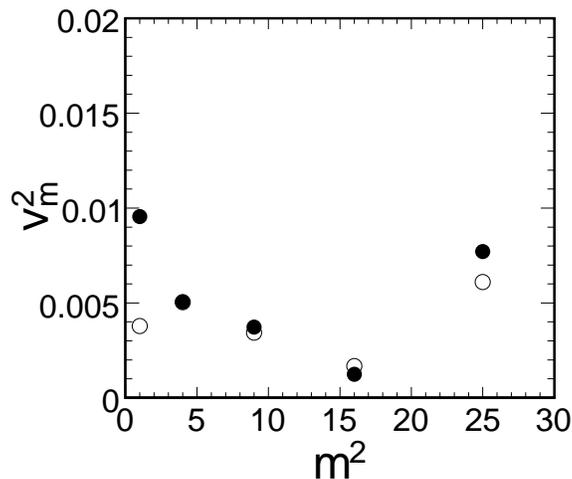,width=8cm,angle=0}}~}
\caption{Plot of $v_m^2$ vs $m^2$ for the case of no jet. The
closed (open) symbols are for with (without) $p_T$ weight.
}\label{fig:nojet}
\end{figure}

The case of only background particles is essentially considered to set
the scale. The background particles are generated from 
HIJING event generators with the option of switching off the jet production and
the hard processes. Therefore, the background particles in HIJING event generators
are only due to the emission of soft particles and no high $p_T$ particles.
Here we do not expect any structure in the flow coefficients
as no jet is present in the data. Ideally, the flow coefficients should
vanish since the particles are distributed uniformly in azimuthal angle.
However, because of fluctuations associated with finite number of
particles, we do obtain small nonzero values. In the limit of infinite
number of background particles, $v_m^2$'s are expected to vanish. The
flow coefficients for one such event without a jet are shown in
Fig(\ref{fig:nojet}). In this figure the value of $p_T$ cut used is  
0.75 GeV and the corresponding number of charged particles are around 60.
One can see that the flow coefficients are indeed
small and there is no systematic variation of these with $m$. Thus a
meaningful linear fit cannot be obtained. The calculations also show that the
flow coefficients systematically decrease as the transverse momentum cut
is decreased ( thus increasing the number of background particles ). This
behavior implies that the non-zero values of the flow coefficients are
indeed due to fluctuations. We have analyzed 3000 such events and
the behavior of the flow coefficients is qualitatively similar to that shown in
Fig(\ref{fig:nojet}). The flow coefficients are generally smaller than
0.005 and 0.015 for constant weight and $p_t$ weighted coefficients
respectively. We don't find any event giving rise to flow coefficients
which would mimic a jet. From this we conclude that these values of
the flow coefficients set the scale for the background. For a meaningful
analysis, the flow coefficients should be significantly larger than these
values.

\subsection{Only Jet particles}
\begin{figure}[htb]
\centerline{~{\psfig{figure=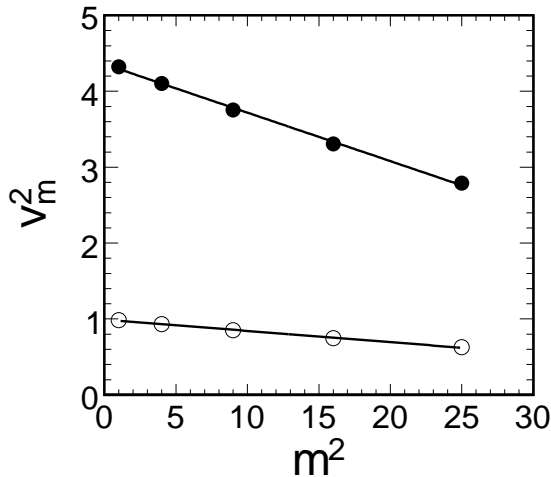,width=8cm,angle=0}}~}
\caption{Plot of $v_m^2$ vs $m^2$ for the case of only jet. The
closed (open) symbols are for with (without) $p_T$ weight.
}\label{fig:onlyjet}
\end{figure}

We now consider the case when there are no background particles but only
 jet particles. This is like a jet produced in $e^+-e^-$ or hadron-hadron
collision. The results obtained for this case are expected to agree very well
with the input. We have put 10 jet particles within
small angle($\Delta \phi$=$\pi$/6) per event. The flow coefficients
$v_m^2$ {\it vs} $m^2$ has been shown in Fig(\ref{fig:onlyjet}) for both with
and without $p_T$ weighted. The values of flow coefficients are large
in comparison with the events having only background particles. The flow
coefficients are nicely fitted with straight line. The intercept on
the y-axis will give the number of jet particles and the transverse
momentum of the jet. One such event is displayed in Fig(\ref{fig:onlyjet}).

\subsection{One Jet with Background}
\begin{figure}[htb]
\begin{minipage}{0.4\textwidth}
\centerline{~{\psfig{figure=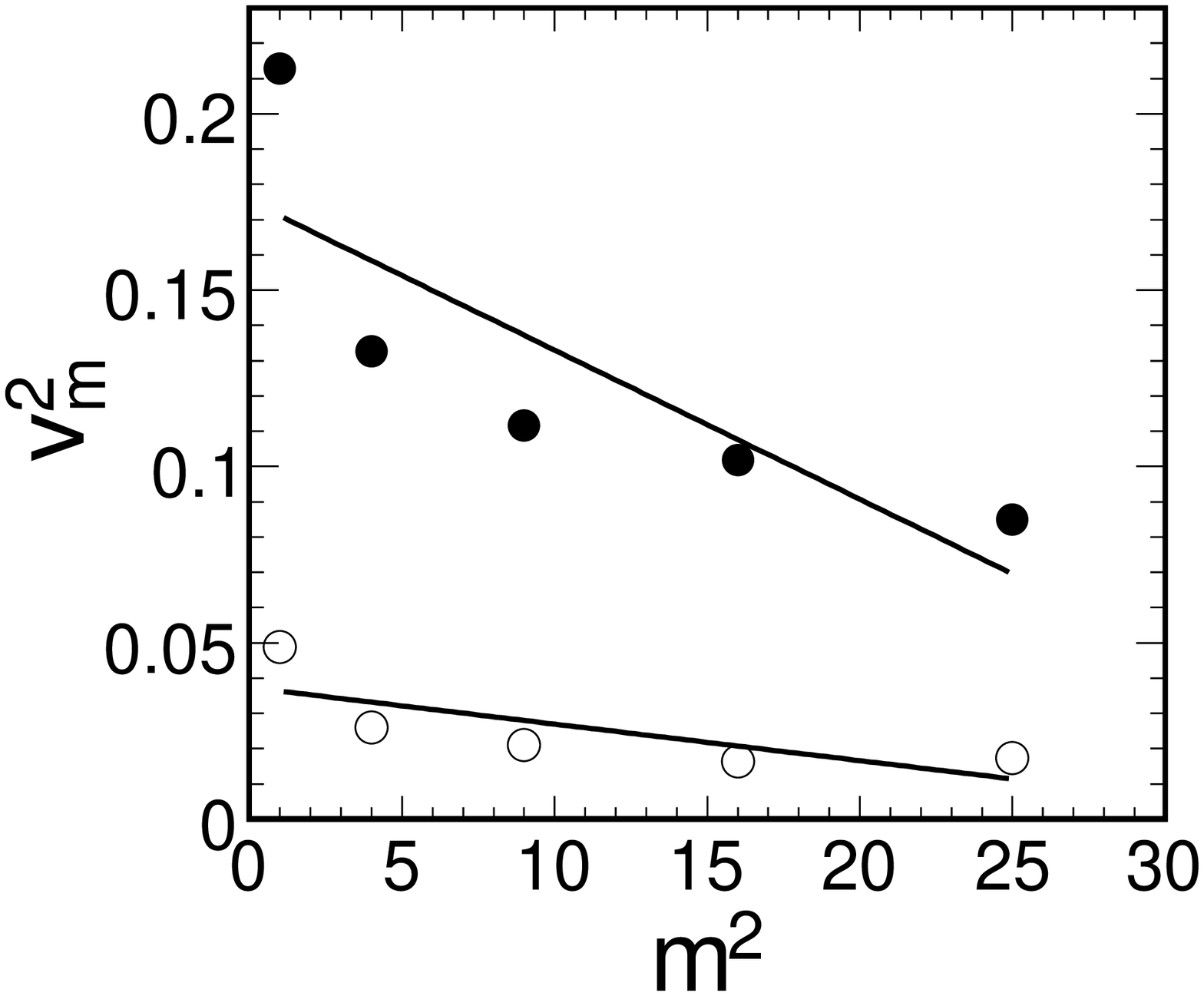,width=6.5cm,angle=0}}~}
\end{minipage}
\hspace*{2cm}
\begin{minipage}{0.4\textwidth}
\centerline{~{\psfig{figure=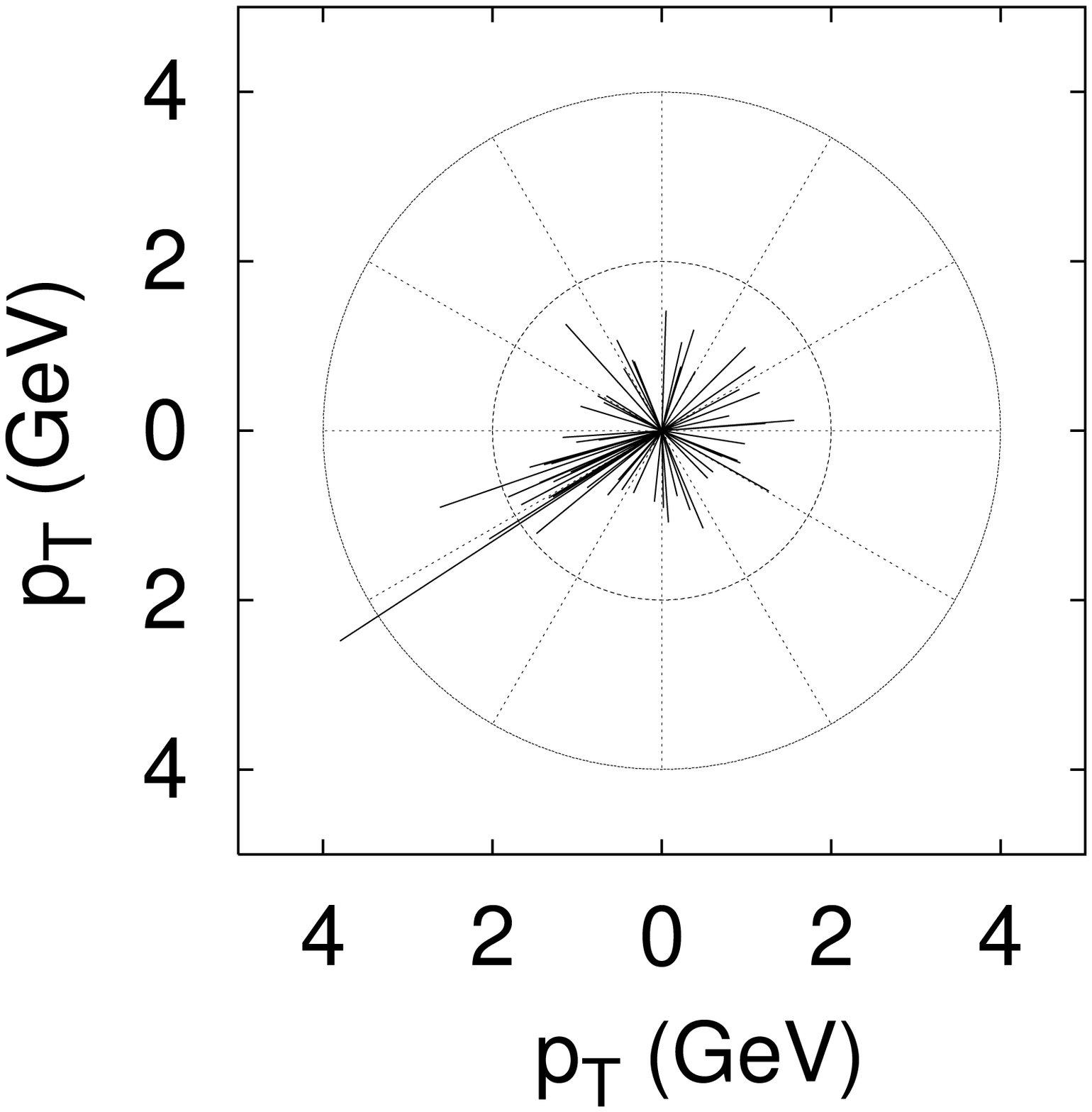,width=6cm,angle=-90}}~}
\end{minipage}
\caption{Plot for $v_m^2$ vs $m^2$ (left figure). The closed (open) symbols
are for with (without) $p_T$ weight. The right figure shows the
transverse momenta of the particles in a 'wagon-wheel' plot. Each particle
is represented by a line in the plot and the length and the direction
give the transverse momentum and azimuthal angle of the particle.}
\label{fig:jetplus}
\end{figure}

We shall now consider the case of one jet with background particles. A
typical plot of $v_m^2$ {\it vs} $m^2$ for $p_T$ cut of 0.75 GeV for
background particles and 10 jet particles is shown in the left panel
of Fig(\ref{fig:jetplus}). The plots for 0.5 and 1 GeV cuts are similar
with $v_m^2$'s being smaller ( larger ) for 0.5 ( 1 ) GeV cut. We find
that the flow coefficients are significantly larger than those obtained
for an event without a jet, implying that the method is likely to work.
Further, a reasonable linear fit to the data can be obtained. Both the
with and without $p_T$-weighted flow coefficients are plotted in the left
panel of Fig(\ref{fig:jetplus}).

Note that $v_m^2$ for $m=2$ is larger than 0.1. As mentioned
earlier, $v_2^2$ for dynamical elliptic flow is expected to be 0.03
or less over a wide range of $p_T$ values \cite{starflow1}. 
Since the observed flow coefficient is
order of magnitude larger than the expected dynamical flow value, we
can definitely conclude that the analysis can be done even in the
presence of non-zero elliptic flow. This is not to say that the
dynamical collective flows cannot be determined by other means.

The right figure in Fig(\ref{fig:jetplus}) shows the 'wagon-wheel'
plot of the same event. In this plot each line represents a particle in
the event and the direction and the length of the line represents the
azimuthal angle and the magnitude of transverse momentum of the particle
respectively.
One can clearly observe a cluster of particles in a small range of azimuthal
angle near $\phi=7\pi/6$. These set of particles are responsible for the peculiar behavior of
the flow coefficients seen in the left panel and these constitute a jet.
In addition to this cluster, one can also observe few other clusters of
fewer particles as well. These are essentially due to statistical
fluctuations in the background. These fluctuations are responsible for
the deviation from the expected linear behavior of the flow coefficients. However,
one should note that in spite of these clusters, the linear pattern
survives.

\begin{table*}
\caption{The extracted values of number of jet particles, jet $p_T$ in GeV,
opening angle in radians and error in these quantities. The errors are due to
the error in obtaining slope and intercept from a linear fit to the
square of the flow coefficients. The average over 3000 events is shown.
The results for $p_T$ cuts of 0.5, 0.75 and 1.0 GeV are shown. The
calculations are done for the jet opening angles of $\pi/6$ and
$\pi/4$ radians. The computations are for 10 jet particles and
jet $p_T$ of 18.26 GeV.}
\vspace*{0.2cm}
\label{Table:jetplus}
\begin{tabular}{|l|cccc|}
\hline
Input \# of particles, jet $p_T$, & \multicolumn{4}{c|}{extracted values} \\
$p_T$ cut and $\Delta \phi $ & \# of particles  & jet $p_T$ & $\Delta
\phi $  & $\Delta \phi $  \\
 &  &  & 
(with $p_T$ weight) & (without $p_T$ weight) \\
\hline
10, 18.26, 1.00,  $\pi/6$ & $ 10.52 \pm 1.37$ & $18.17 \pm 1.75$ & $ 0.43 \pm 0.09$ &
$ 0.46 \pm 0.08$ \\
10, 18.26, 0.75,  $\pi/6$ & $ 11.60 \pm 2.29$ & $18.17 \pm 2.47$ & $ 0.46 \pm 0.07$ & $
0.57 \pm 0.008$ \\
10, 18.26, 0.50,  $\pi/6$ & $ 14.11 \pm 3.46$ & $18.14 \pm 3.11$ & $ 0.56 \pm 0.02$ &
$ 0.70 \pm 0.10$ \\
10, 18.26, 1.00,  $\pi/4$ & $ 10.31 \pm 1.27$ & $18.17 \pm 1.65$ & $ 0.54 \pm 0.07$ &
$ 0.53 \pm 0.08$ \\
10, 18.26, 0.75,  $\pi/4$ & $ 11.44 \pm 2.14$ & $18.17 \pm 2.31$ & $ 0.53 \pm 0.08$ &
$ 0.57 \pm 0.04$ \\
10, 18.26, 0.50,  $\pi/4$ & $ 14.05 \pm 3.30$ & $18.18 \pm 2.85$ & $ 0.54 \pm 0.08$ &
$ 0.77 \pm 0.16$ \\
10, 18.26, 1.00,  $\pi/8$ & $ 10.56 \pm 1.41$ & $18.17 \pm 1.80$ & $ 0.39 \pm 0.07$ &
$ 0.44 \pm 0.06$ \\
10, 18.26, 0.75,  $\pi/8$ & $ 11.63 \pm 2.36$ & $18.17 \pm 2.55$ & $ 0.45 \pm 0.04$ &
$ 0.59 \pm 0.03$ \\
10, 18.26, 0.50,  $\pi/8$ & $ 14.23 \pm 3.55$ & $18.19 \pm 3.18$ & $ 0.51 \pm 0.03$ &
$ 0.97 \pm 0.37$ \\
\hline
\end{tabular}
\end{table*}
The results obtained after analyzing over 3000 events are summarized
in Table - \ref{Table:jetplus}. The simulation is done for three different
jet opening angles ( $\Delta \phi$ ) and three transverse momentum cuts
are employed.  The average of the extracted number of particles, jet
transverse momentum and jet opening angle as well as the average estimated
error is shown. Note that the error quoted is due to the uncertainty
in the determination of the slope and intercept from $\chi^2$ fitting.
Following conclusions can be drawn from these results.
\begin{itemize}
\item The extracted values of the opening angle $\Delta \phi$ are
correlated with the corresponding input values. However, these are
smaller for large opening angles. The errors in the extracted values
are also large ( between 10 and 30 \% ).
The extracted opening angle values for without $p_T$-weight case are
systematically higher for $p_T$ cut of 0.5 GeV.
The agreement is better for
smaller opening angles. The failure at the larger opening angle can
be attributed to the failure of the expansion of the Bessel function
in powers of $\Delta \phi^2 m^2$ for larger $m$'s. One generally extracts
larger slope and therefore larger opening angle if one uses fewer
values of $m$ for fitting.
\item  For smaller $p_T$ cut the value of extracted opening angle is large
and the error in it also equally large for without $p_T$ weighted values.
\item Extracted values of number of jet particles agree better for
larger $p_T$ cut. For the $p_T$ cut of 0.5 GeV the extracted number
of jet particles is systematically larger by 40\%. For 1 GeV cut, the
agreement is very good. The error in the extracted values decreases
almost linearly with $p_T$ cut.
\item The extracted values of jet transverse momentum, on the other
hand, agrees very well ( within 5\% ) with the input value.
\end{itemize}

The reason for the $p_T$-weighted analysis working better than the
constant weight can be  understood as follows. Most of the
background particles are having  transverse momenta smaller than
1 GeV or so. Thus when one computes the $p_T$-weighted flow coefficients,
the importance of the background particles in the flow coefficients is
de-emphasized and large transverse momentum particles are given larger
weight. Hence, to some extent, $p_T$ weighting plays the same role as
that of transverse momentum cut. As a result the extracted $p_T$ of the
jet does not appear to be sensitive to the transverse momentum cut
applied in the analysis.

\begin{figure}[htb]
\centerline{~{\psfig{figure=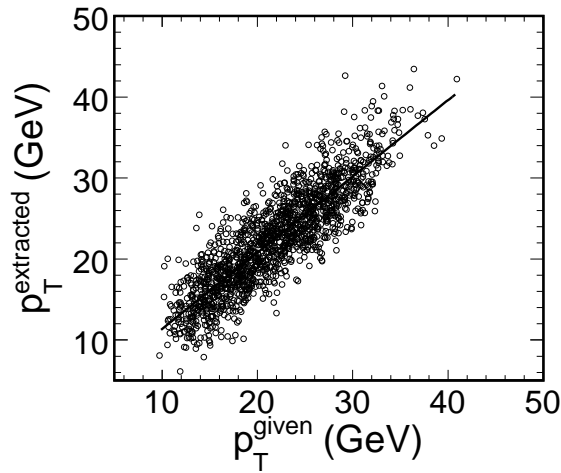,width=8cm,angle=0}}~}
\caption{Plot of extracted jet transverse momentum {\it vs} input
transverse momentum. The graph shows a linear correlation between
the two.}
\label{fig:pT_corr}
\end{figure}

In our simulation of the jet event, the number of jet particles is
fixed but the jet transverse momentum is not. This is because the
transverse momentum of the jet particles is assigned statistically
according to the transverse momentum distribution function $P(p_T)$.
Thus, although the jet transverse momentum is, on the average, given
by the product of the average transverse momentum $<p_T>$ times
the number of jet particles, the jet transverse momentum fluctuates,
from event to event, about this number. This means that for a set of
events with fixed number of jet particles, the jet transverse momentum
is distributed over a range. If we consider the jets with different
number of particles in a jet, we essentially generate a number of events
with a broad distribution of jet transverse momenta. We can then look
at the correlation between the input and extracted jet transverse
momenta. Such a  correlation between extracted and input jet transverse
momentum is displayed in Fig(\ref{fig:pT_corr}). One can fit a nice
straight line with the slope close to unity and the intercept close to
zero, although the $\chi^2$ is rather large at 5.6.

\subsection{Two Jets with Background}
\begin{figure}[htb]
\begin{minipage}{0.4\textwidth}
\centerline{~{\psfig{figure=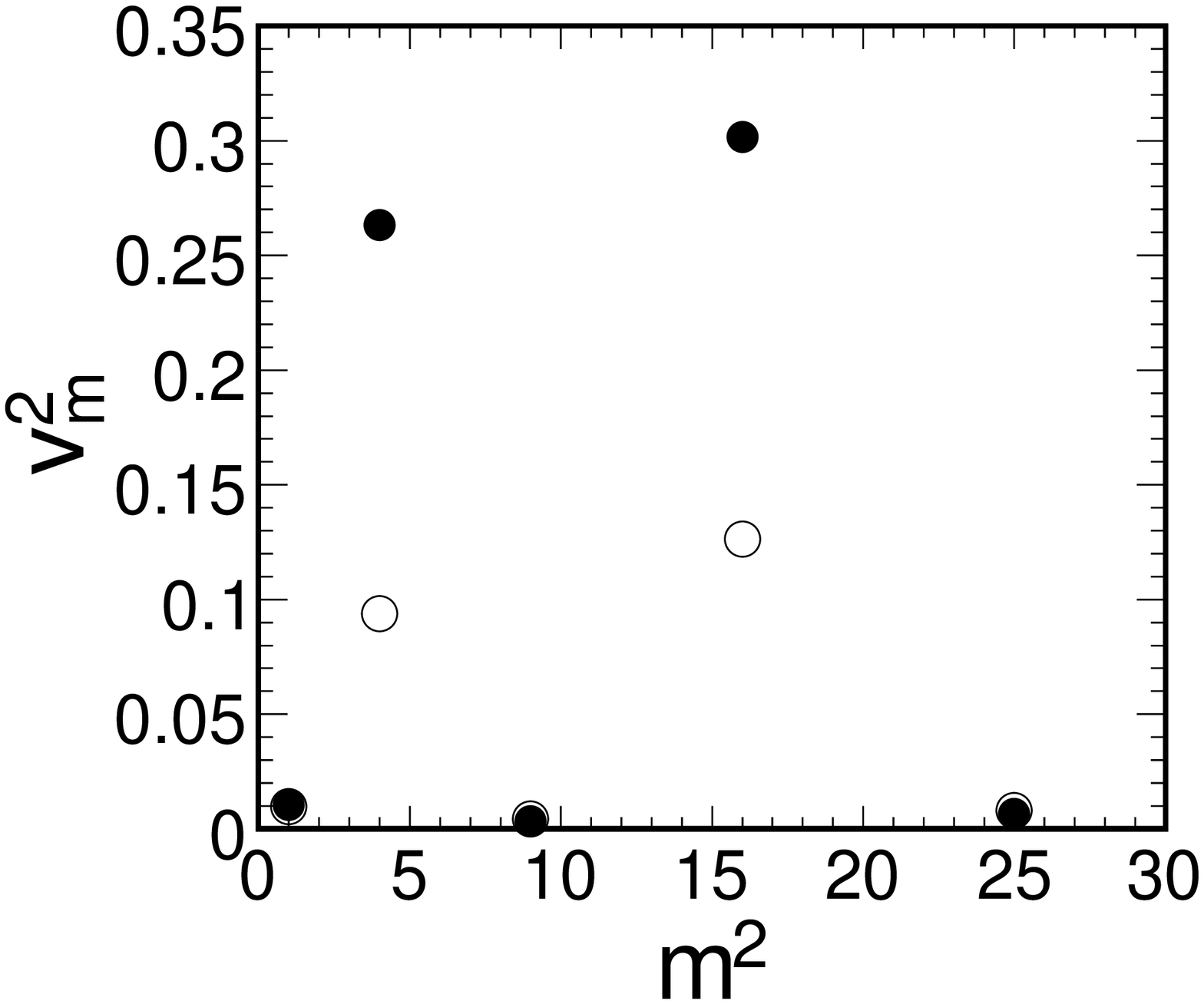,width=6.5cm,angle=0}}~}
\end{minipage}
\hspace*{2cm}
\begin{minipage}{0.4\textwidth}
\centerline{~{\psfig{figure=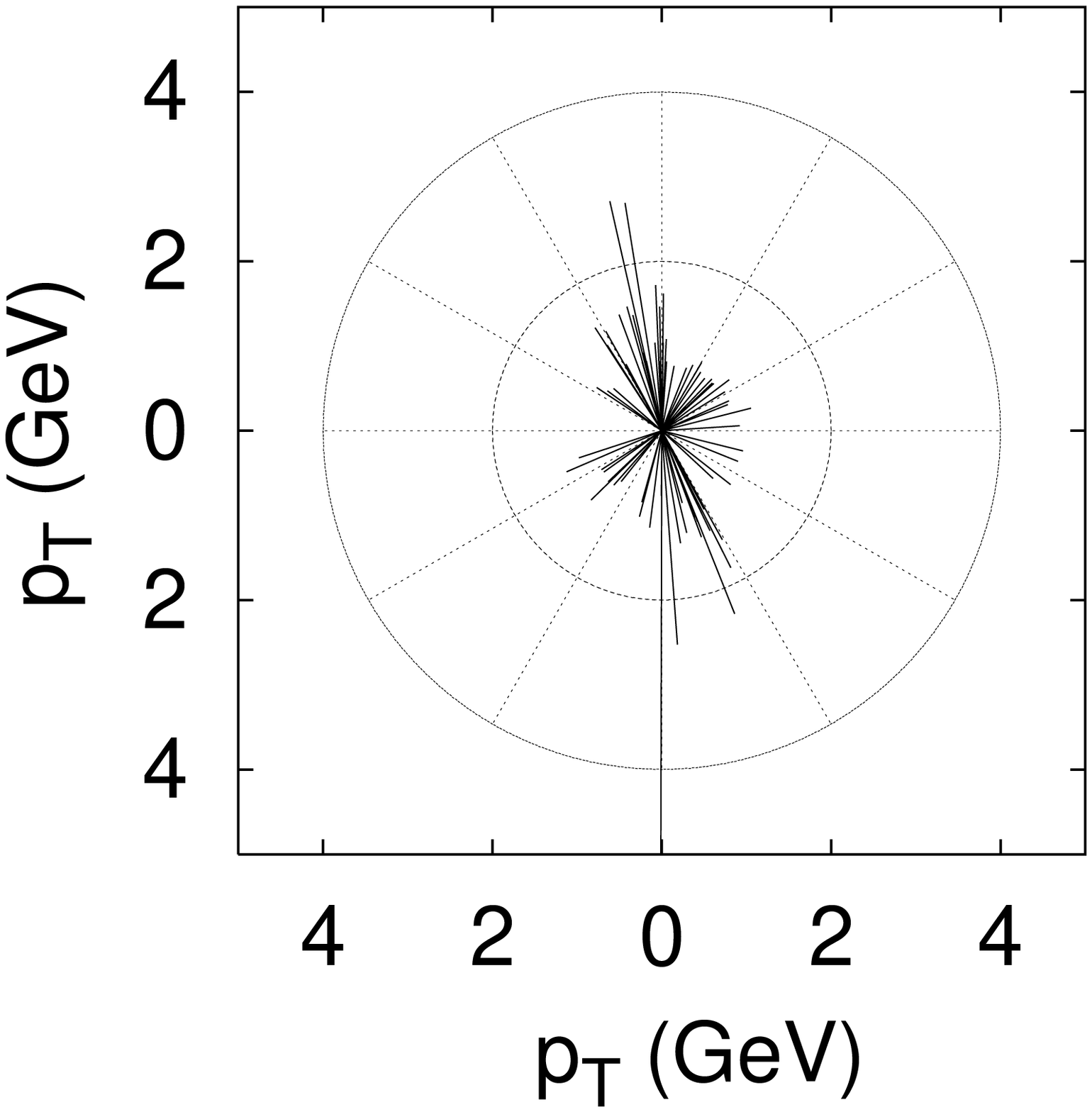,width=6cm,angle=-90}}~}
\end{minipage}
\caption{Plot of $v_m^2$ vs $m^2$ (left figure).
The closed (open) symbols are for with (without) $p_T$ weight for $p_T$ cut of 0.75 GeV.
The figure on the right shows the transverse momenta of the particles
in a 'wagon-wheel' plot. Two back-to-back jets can be identified in the figure.
}
\label{fig:twojetplus}
\end{figure}

We have noted earlier that when we have two back-to-back jets,
the flow coefficients show a distinct odd-even effect with the
odd $v_m^2$'s being very small in comparison with those for even
$m$. The suppression of the odd flow coefficients is maximum when
the two jets have same opening angle and equal number of jet particles.
We have also investigated the effect of varying the opening angle and
number of particles of one of the jets. This, in some sense, is
equivalent to partial quenching of one of the jets. The results of
the investigation are shown in Fig (\ref{fig:twojetplus}). The
closed symbols of left figure show the $v_m^2$ vs $m^2$ having
different opening angle of both the jets. The open symbols are for
both the jets having same opening angle.

When the number of jet particles in one of the jets is reduced, the
contribution of that jet to the flow coefficients decreases and
naturally the plot starts looking like a single jet. The
corresponding transverse momenta of the jet particles are shown
in the 'wagon-wheel' plot in Fig (\ref{fig:twojetplus}).

\begin{figure}[htb]
\begin{minipage}{0.4\textwidth}
\centerline{~{\psfig{figure=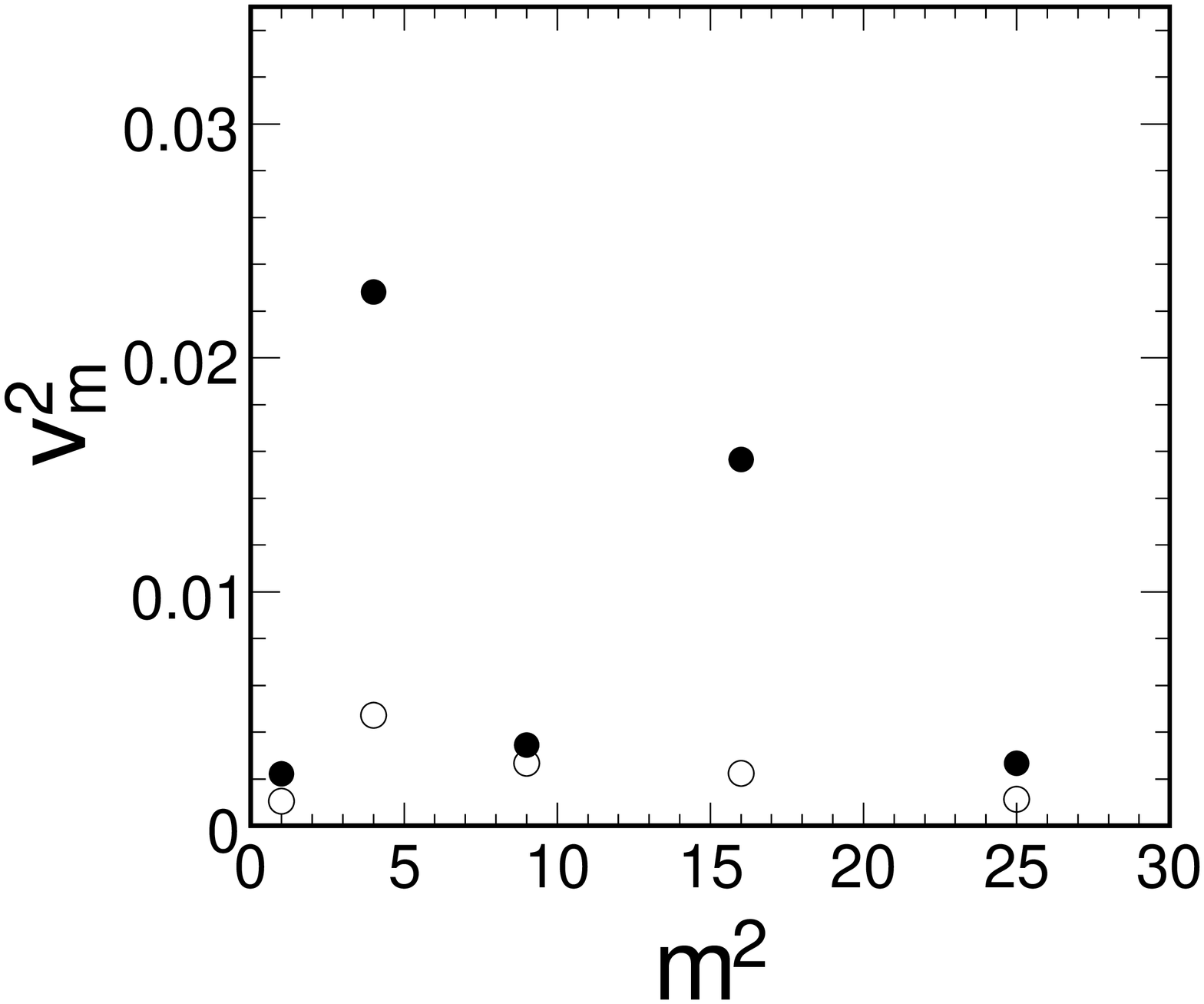,width=5.5cm,angle=-90}}~}
\end{minipage}
\hspace*{2cm}
\begin{minipage}{0.4\textwidth}
\centerline{~{\psfig{figure=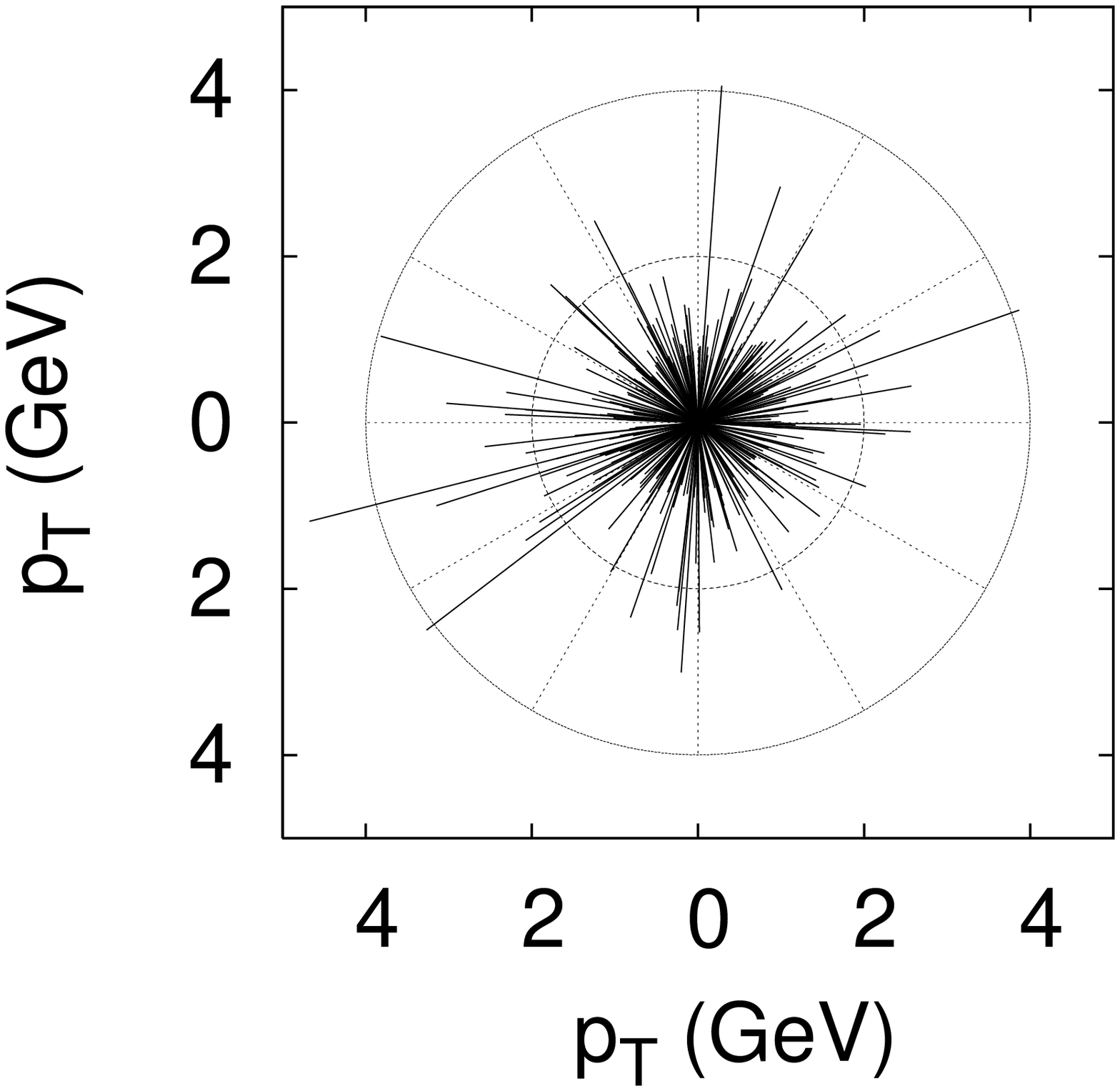,width=6cm,angle=-90}}~}
\end{minipage}
\caption{Plot of $v_m^2$ vs $m^2$ (left figure). 
The closed (open) symbols are for with (without) $p_T$ weight for $p_T =0.75$ GeV cut.
The figure on the right shows the transverse momenta of the particles
in a 'wagon-wheel' plot. Two jets can be identified in the figure.
}
\label{fig:Hijingtwojet}
\end{figure}

A few comments are in order at this stage. First, one does not really get 
back-to-back jets even in hadron-hadron collision because the partons have 
internal motion within a hadron and sometimes a scattered parton may emit 
another hard parton, thus producing a three-jet structure. In the first case, 
the angle between the two jets would not be $180 ^o$ but close to it. In 
that case, the odd flow coefficients will not vanish but would still be small. 
In the second case, we really do not have two jets so we should not expect 
to have the typical odd-even effect of two jets event. In case of 
nucleus-nucleus collision, one of the jets may be quenched or particles 
of one of the jets may scatter from the background. This would mean that 
the opening angle of this jet would be larger or there would be fewer 
particles in this jet. This means that as this effect becomes stronger, 
the structure of the event would go over from two jets event, with typical 
odd-even effect to single jet structure. 
In Fig(\ref{fig:Hijingtwojet}), we have shown two jets event from 
HIJING event generators by switching on the jets. In this figure, the flow coefficient
show a distinct odd-even structure in the plot of
$v_m^2$ vs $m^2$, as we discussed above. In the next section we will 
discuss the events in HIJING by turning on the jet production in detail.

\begin{figure}[htb]
\begin{minipage}{0.4\textwidth}
\centerline{~{\psfig{figure=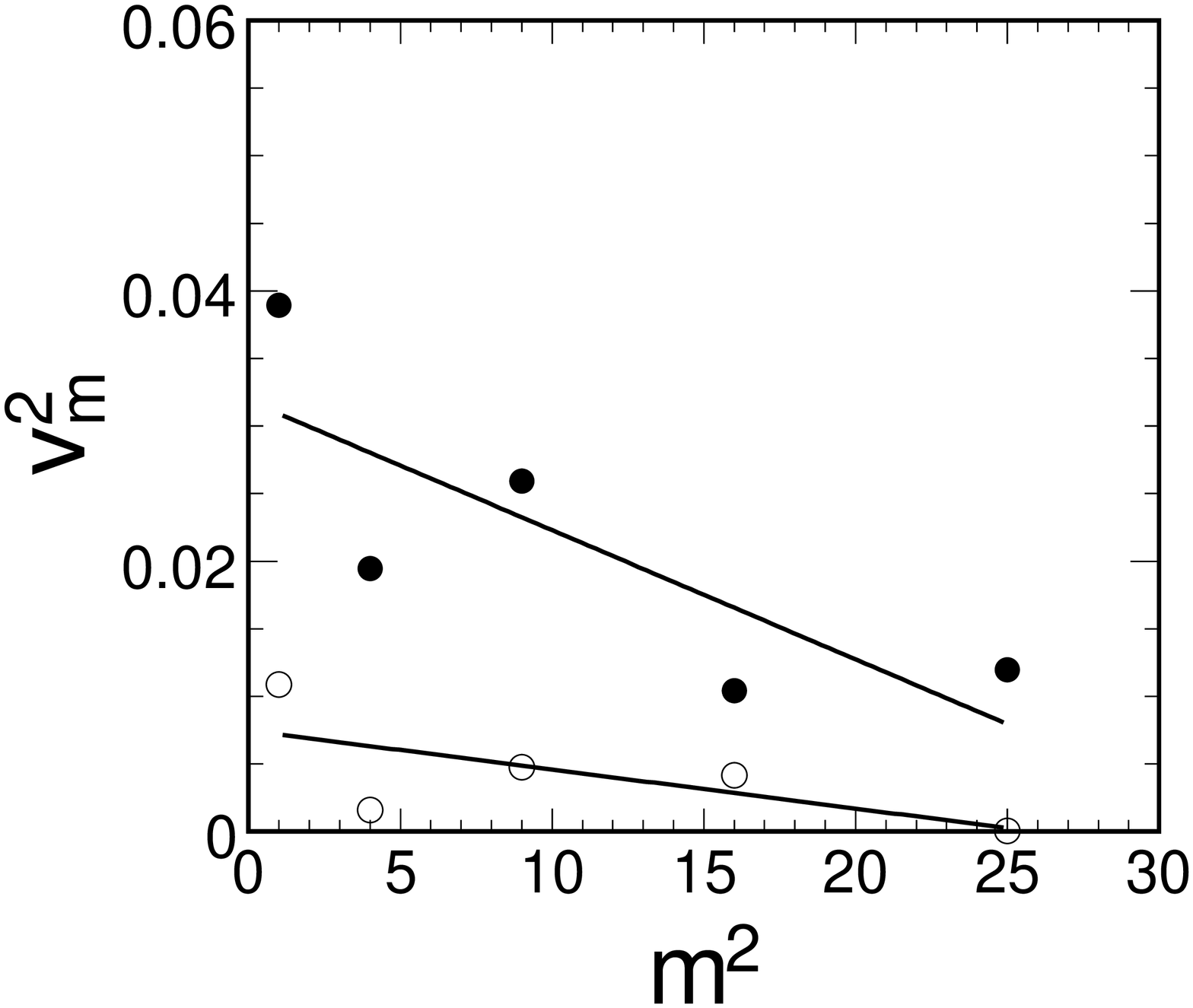,width=5.5cm,angle=-90}}~}
\end{minipage}
\hspace*{2cm}
\begin{minipage}{0.4\textwidth}
\centerline{~{\psfig{figure=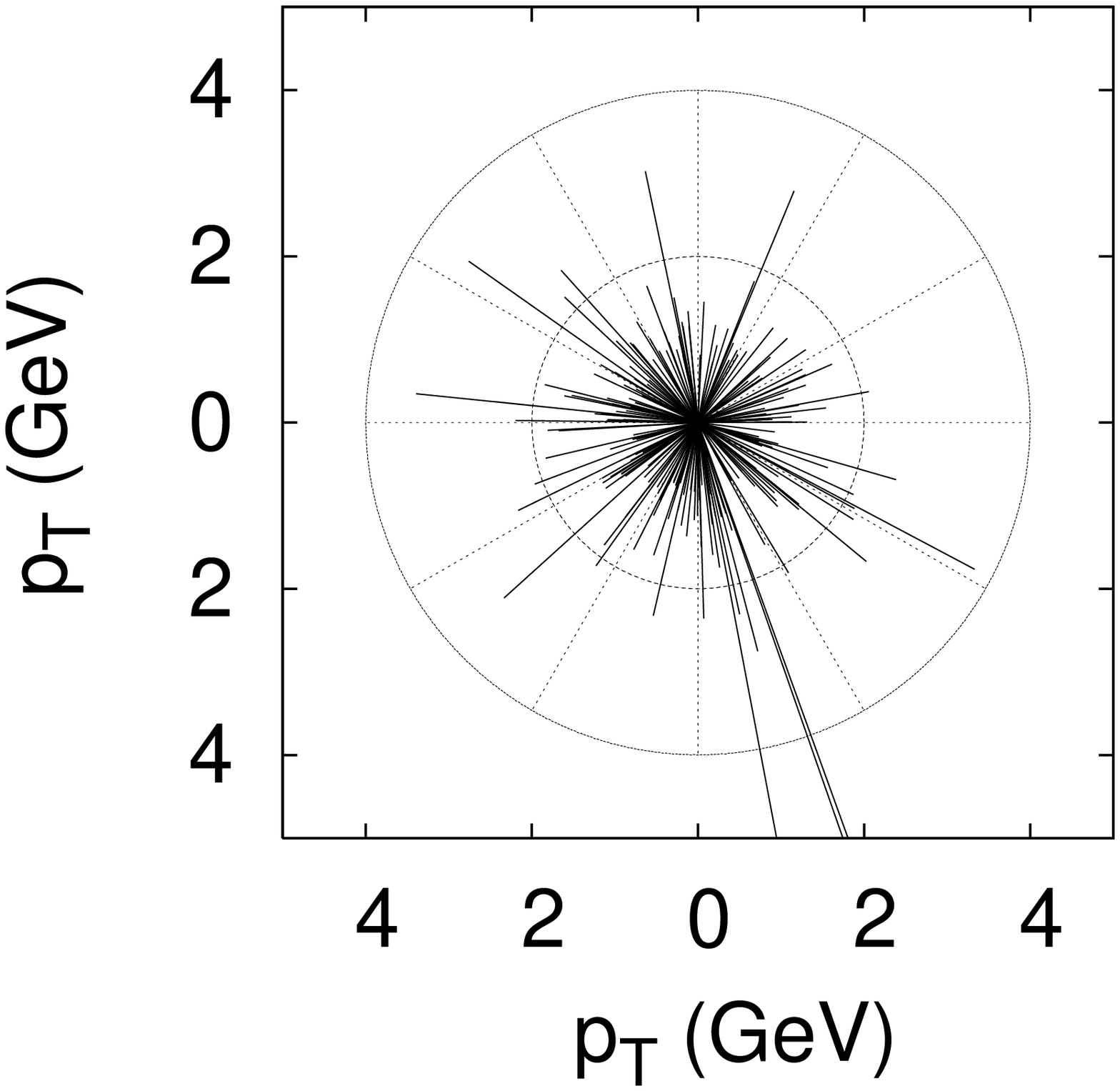,width=6cm,angle=-90}}~}
\end{minipage}
\caption{Plot of $v_m^2$ vs $m^2$ (left figure).
The closed (open) symbols are for with (without) $p_T$ weight for $p_T =0.75$ GeV cut.
The figure on the right shows the transverse momenta of the particles
in a 'wagon-wheel' plot. One jets can be identified in the figure. The extracted
values of jet particles are $17.7 \pm 4.4$, jet $p_T=29.6 \pm 7.3$  GeV, $\Delta \phi= 
0.72\pm 0.34$( with $p_T$ weight) and $\Delta \phi= 0.70 \pm 0.32$ ( without
$p_T$ weight).
}
\label{fig:Hijingonejet}
\end{figure}

\begin{figure}[htb]
\begin{minipage}{0.4\textwidth}
\centerline{~{\psfig{figure=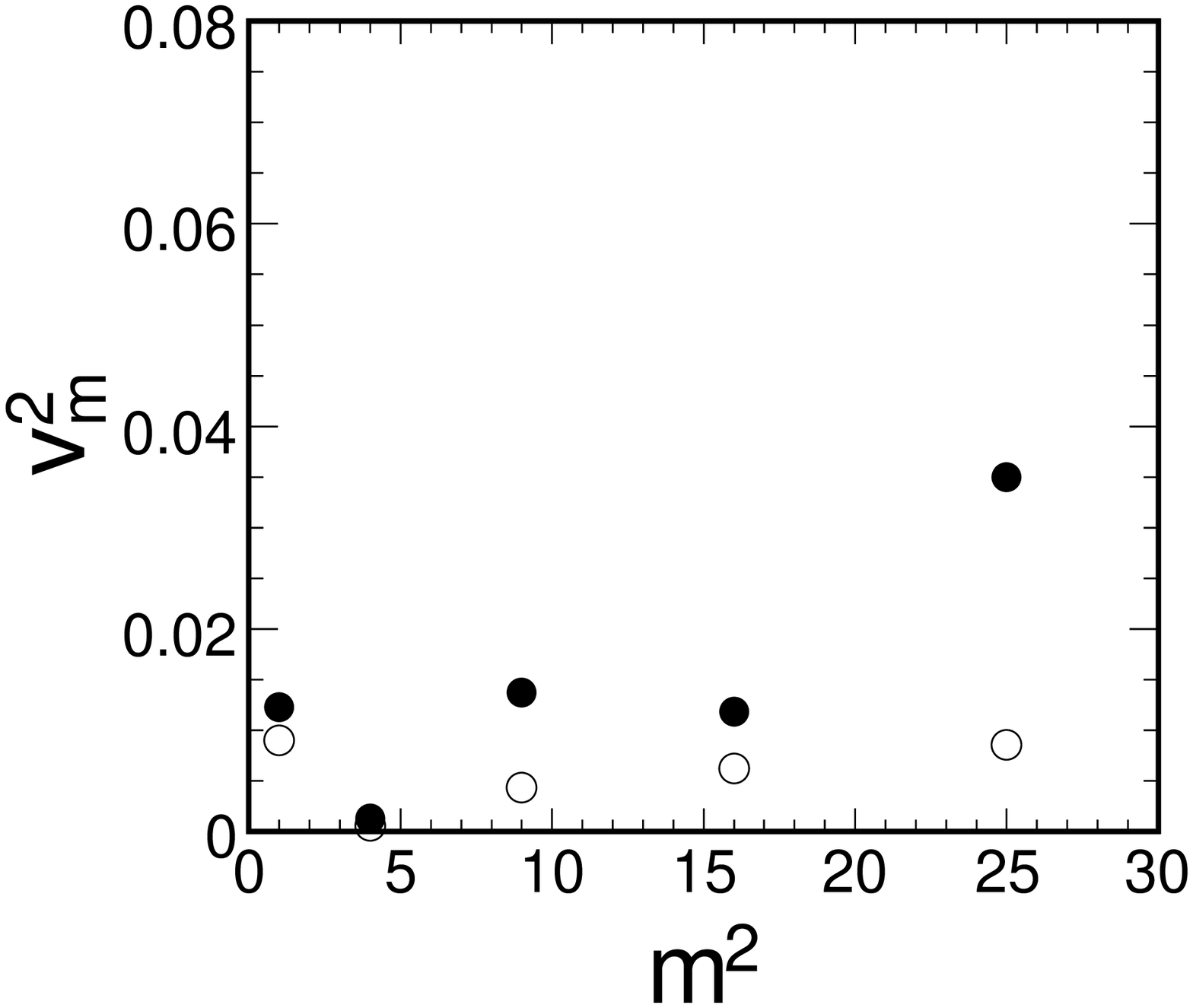,width=7cm,angle=0}}~}
\end{minipage}
\hspace*{2cm}
\begin{minipage}{0.4\textwidth}
\centerline{~{\psfig{figure=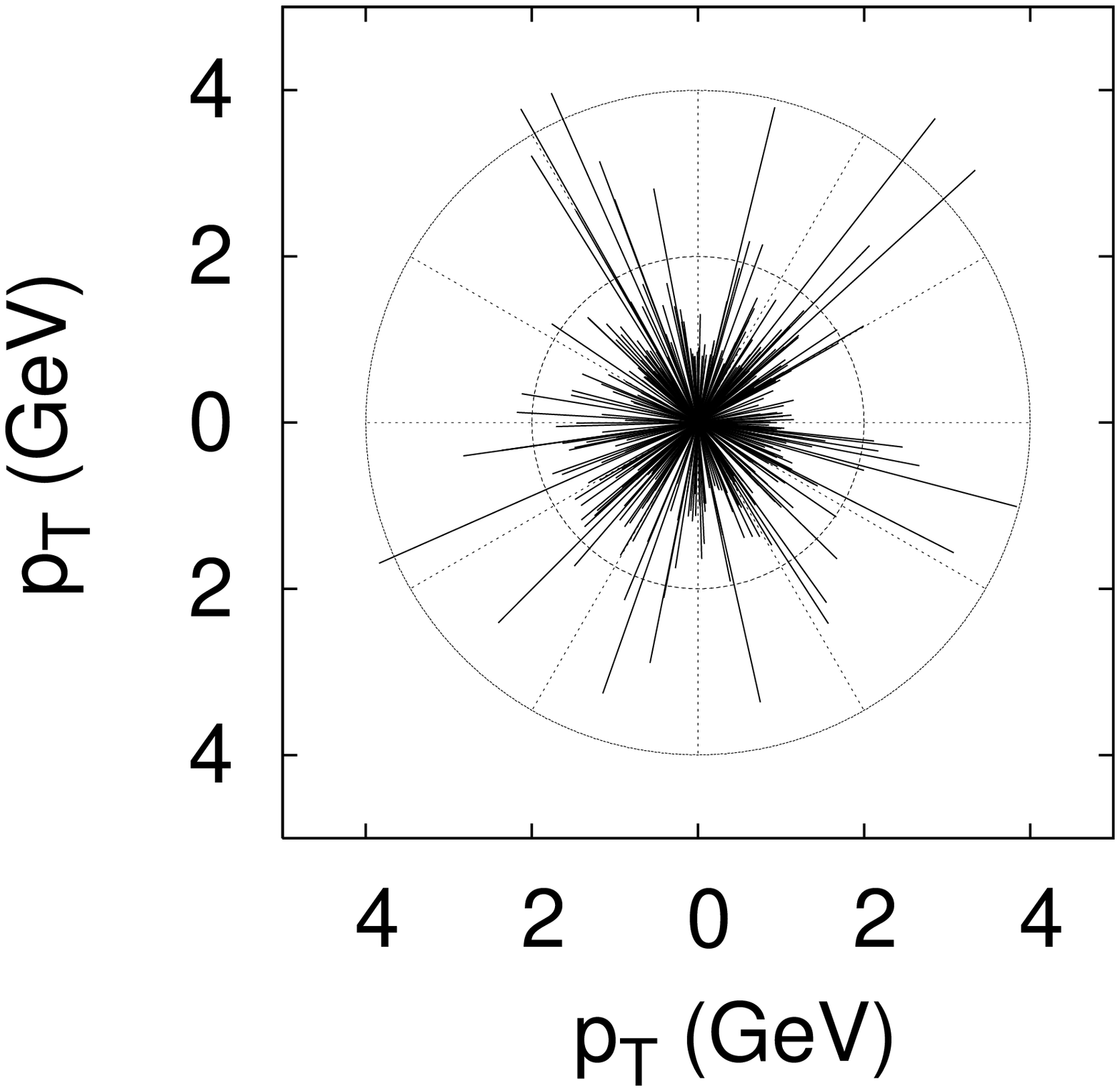,width=6cm,angle=-90}}~}
\end{minipage}
\caption{Plot of $v_m^2$ vs $m^2$ (left figure).
The closed (open) symbols are for with (without) $p_T$ weight for $p_T =0.75$ GeV cut.
The figure on the right shows the transverse momenta of the particles
in a 'wagon-wheel' plot. This event shows no-jet by our method,
though jets are present in this event, which is generated by HIJING.
}
\label{fig:Hijingnojet}
\end{figure}

\subsection{HIJING Events with Jets}

We now discuss the results for HIJING events with jets present. As 
mentioned earlier, high energy jets with large numbers of jet particles, 
which correspond to ( relatively ) hard collision of partons are rare.  
We are using HIJING parameters which allow maximum number of hard 
scatterings per nucleon-nucleon collision. It is not clear if this number is 
realistic but it does help in generating events having large energy jets having
large number of jet particles. Of course most of the jets are mini-jets having 
few jet particles and our method is not expected to detect such mini-jets. 

When an event has a single high energy jet with large number of particles 
( and rest of the particles being produced by low energy jets and other 
background particles ), the flow coefficients are expected to have a 
typical behavior like those events considered above in Section 3.3 above. 
Further, when there are two almost back-to-back jets, the event would be 
similar to those discussed in Section 3.4 above. In other cases, the 
event is expected to look like the background only events discussed in 
Section 3.1 above. This is because, many low energy jets will have the 
azimuthal angle distribution of particles similar to that of background 
particles. Thus, by studying the flow coefficients we would be able to 
classify the HIJING events into three categories, namely, one jet events 
which have large flow coefficients, two jet events having oscillating 
flow coefficients and rest which have small flow coefficients which are 
almost random, not having any pattern.

We have analysed a few thousand HIJING events with jets switch on and we 
find that 50 to 60 \% of the events can be classified as one or two jet events. 
The wagon-wheel plot as well as the plot of flow coefficients for one of 
the HIJING event identified as a single jet event is shown in 
Fig(\ref{fig:Hijingonejet}). Also similar plot for a typical event which 
has been classified as no-jet event by our method is displayed in 
Fig(\ref{fig:Hijingnojet}). There is a clear-cut correlation between the 
plot of flow coefficients ( which indicates whether the event would have 
jet structure or not ) and the wagon wheel plot ( which shows a jet on 
visual inspection ).

\section{Conclusions}
We have explored the possibility of identifying and characterizing the
jet structure in a relativistic heavy ion collisions. The method exploits
the fact that if the event has sufficiently large number of particles
emitted in a narrow cone in azimuthal angle, the flow coefficients for
such an event are abnormally large. Further, we have shown that in such
a case, there is a linear relation between the square of the flow
coefficients $v_m$ with $m^2$ and using this relation it is possible
to estimate the number of jet particles, the jet opening angle and the
jet transverse momentum. For the last quantity one has to compute the
$p_T$ weighted flow coefficients. We have applied the method to simulated
data having zero, one and two jets. We find that these three cases can
be distinguished from the pattern of the flow coefficients. For the events
with no jet, the flow coefficients are small. For one jet events, the flow
coefficients are large and show the linear behavior discussed above. In
case of the two jets, the odd and even flow coefficients fluctuate with
the odd coefficients being small. We believe that from the observed
behavior of the flow coefficients in an event, it would be possible
to identify events in which one of the jet suffers quenching.
We feel that this method can be used in LHC experiments to isolate collision
events having one and two jets and possibly extract the properties of jets.


\begin{thebibliography}{99}
\bibitem{defn} An alternate expansion is
$P(\phi) = v_0 + \sum_{m=1,\infty} ( a_m \cos(\phi) + b_m \sin(\phi))$
with $v_m = \sqrt{ a^2 + b^2}$ and $\phi_m = \tan^{-1}a_m/b_m$.
\end{thebibliography}
\end{document}